# The Instantaneous Breit Equation with an Application to Charmonium

Walter S. Jaronski

Department of Physics, Radford University, Radford, Virginia 24142, United States
Email: wjaronsk@radford.edu

**Abstract.** The solution of the Breit equation with an instantaneous potential for the case of two spin-½ particles in a pseudoscalar bound state is considered. This is then applied to charmonium using a potential of the Cornell type. The masses of the two $J^P = 0^-$ states below charm threshold (the $\eta_c$ and $\eta_c'$) are calculated in this model. We allow different mixtures of the Lorentz nature (vector or scalar) of the linear confining term and investigate the effect of these on the bound-state energies. Some general comments are made on the issue of how the bound nature of these states depends on the vector-scalar mix.

**Keywords:** Breit equation, two-body Dirac equation, linear potential, Lorentz vector and scalar potentials, quarkonium, charmonium

## 1  Introduction

We are interested in what should be a very basic problem: the relativistic treatment of the bound state of a pair of interacting particles. Contrary to the nonrelativistic case, where the solution is routine, there are many difficult issues once special relativity is included. This is true in both classical physics and quantum theory. In both cases, the lack of a simple center of mass prevents a reduction to an equivalent one-body problem. In both cases, the requirement of including time coordinates for each particle presents a mathematical complication as well as an obvious difficulty in interpretation. In the classical case, one must also contend with the Currie-Jordan-Sudarshan "no-go" theorem[1], which asserts that Hamiltonian mechanics and relativistic covariance preclude the possibility of interaction. This is no longer an issue in the quantum case, but we then have the more serious issue of particle creation and annihilation, which prevents a theory of a fixed number of particles (two, in this case) from being a complete solution to the problem. One is then forced to abandon the hope that a standard quantum-mechanical two-body equation will be rigorously correct and must treat the problem using the methods of quantum field theory.

This in undoubtedly fundamentally correct and there is a generally recognized approach to the field theory of bound states – the Bethe-Salpeter equation.[2] But this approach itself is beset with difficulties, both mathematical complications and issues of interpretation. Quasipotential approaches[3], which are approximations off the Bethe-Salpeter equation, alleviate some, but not all, of these issues (or at least not completely). There are also many different quasipotential approaches[4] with various levels of difficulty in implementation and interpretation, which may have different results in a specific calculation.[5] One might then seek a two-body equation which allows a routine quantum-mechanical interpretation, but with an acknowledgement of its limitations. It is recognized that the equation resides in the two-particle sector of Fock space[6] and that corrections due to mixing with other sectors of the space must be considered for greater accuracy. One obvious candidate for such an equation is that of Breit.[7] We are restricting ourselves now to spin-½ particles since they are the fundamental matter particles. The Breit equation includes a Dirac Hamiltonian for each of the particles and an interaction potential. It thus has the form

$$(H_1 + H_2 + V)\Psi = E\Psi, \tag{1}$$

where $H_i = \vec{\alpha}_i \cdot \vec{p}_i + \beta_i m_i$ for particles "1" and "2" with masses $m_1$ and $m_2$, and $\vec{\alpha}$ and $\beta$ are the standard Dirac matrices. We will take the potential $V$ to be instantaneous and central: $V = V(r)$ where $r$ is the magnitude of the relative coordinate. Breit was focused on the electromagnetic interaction between two electrons for which $V = e^2/r$ and included a term

$$V_B = -\frac{1}{2}V(\vec{\alpha}_1 \cdot \vec{\alpha}_2 + \vec{\alpha}_1 \cdot \hat{r}\vec{\alpha}_2 \cdot \hat{r}) \tag{2}$$

to account for retardation. This term is a quantum version of the classical Darwin Lagrangian; we will ignore the discussion of its adequacy in treating retardation effects. In our instantaneous approximation, this term will not be included; there will be further discussion of this in Section 5 below. We will work in the center of momentum frame and will take $\vec{p}_1 = -\vec{p}_2 \equiv \vec{p}$. Then Eq. (1) becomes

$$[(\vec{\alpha}_1 \cdot \vec{p} + \beta_1 m_1) + (-\vec{\alpha}_2 \cdot \vec{p} + \beta_2 m_2) + V]\Psi = E\Psi. \tag{3}$$

In this equation, $\alpha_1$ and $\alpha_2$ are the same standard 4 x 4 matrix, as are $\beta_1$ and $\beta_2$; the subscripts are added just to remind us that they are acting on particle 1 and particle 2, respectively,

In the following discussion, we will assume, as stated above, that the potential is a function of the magnitude of the relative spatial coordinate only, $V = V(r)$. This violates Lorentz covariance, but it is needed to make the treatment of the equation mathematically tractable. Moreover, an instantaneous potential might be reasonable at the level of the applications we have in mind (primarily heavy quarkonium) and, in any case, Eq. (3) still includes relativistic kinematics and spin, and is therefore an improvement over many other nonrelativistic or semi-relativistic treatments (such as the spinless Salpeter approach[8] in which relativistic kinematics and spin dependence are put in by hand). We acknowledge other valuable approaches[9] to the two-body problem which treat the relative-time issue using Dirac's constraint theory[10], but have chosen to maintain the simplicity of the instantaneous approach.[11]

## 2 Reduction of the equation

Eq. (3) is a two-body Dirac equation for a wave function of 4 x 4 = 16 components. Such a wave function can be written as a 16-component column vector that is the direct product of two one-particle Dirac spinors. For the case of two 4-component Dirac spinors $\psi = \{\psi_\mu\}$ and $\varphi = \{\varphi_\mu\}$, $\mu = 1, \ldots, 4$, we have

$$\psi \otimes \varphi = \begin{pmatrix} \psi_1 \varphi_1 \\ \psi_1 \varphi_2 \\ \vdots \\ \psi_4 \varphi_3 \\ \psi_4 \varphi_4 \end{pmatrix}. \tag{4}$$

But it is more convenient to arrange it as a 4 x 4 matrix; we can form this by using the following matrix multiplication

$$\begin{pmatrix} \psi_1 \\ \psi_2 \\ \psi_3 \\ \psi_4 \end{pmatrix} (\varphi_1 \quad \varphi_2 \quad \varphi_3 \quad \varphi_4) = \begin{pmatrix} \psi_1\varphi_1 & \psi_1\varphi_2 & \psi_1\varphi_3 & \psi_1\varphi_2 \\ \psi_2\varphi_1 & \psi_2\varphi_2 & \psi_2\varphi_3 & \psi_2\varphi_4 \\ \psi_3\varphi_1 & \psi_3\varphi_2 & \psi_3\varphi_3 & \psi_3\varphi_4 \\ \psi_4\varphi_1 & \psi_4\varphi_2 & \psi_4\varphi_3 & \psi_4\varphi_4 \end{pmatrix} \equiv \Psi_{\mu\nu}. \tag{5}$$

This has the same terms as in the linear direct product; it is simply a different arrangement of these terms. The following discussion will not be limited to the product case, but this product form shows the breakdown of our "matrix wave function" into quadrants, with the upper left quadrant being the large-large components, the upper right quadrant being the large-small components, the lower left quadrant being the small-large components, and the lower right quadrant being the small-small components (in the nonrelativistic limit). The general expression for $\Psi$ is

$$\Psi = \begin{pmatrix} \Psi_{11} & \Psi_{12} & \Psi_{13} & \Psi_{14} \\ \Psi_{21} & \Psi_{22} & \Psi_{23} & \Psi_{24} \\ \Psi_{31} & \Psi_{32} & \Psi_{33} & \Psi_{34} \\ \Psi_{41} & \Psi_{42} & \Psi_{43} & \Psi_{44} \end{pmatrix}. \tag{6}$$

We must then consider the action of operators on a wave function in this matrix form. If A is a matrix operator acting on particle 1 and B is a matrix operator acting on particle 2, the joint action of A and B on the matrix wave function $\Psi$ is given by

$$\Psi' = A\Psi B^T, \quad i.e. \ \Psi'_{\mu\nu} = A_{\mu\rho}\Psi_{\rho\sigma}B^T_{\sigma\nu} = A_{\mu\rho}\Psi_{\rho\sigma}B_{\nu\sigma}. \tag{7}$$

This is due to the specific form of the matrix wavefunction – note that in the product form the wave function for particle 1 is along the columns in $\Psi$ and the wave function for particle 2 is along the rows. Eq. (7) shows a distinct advantage of the matrix form of the wave function. In this form the action of product operators is the matrix product of 4 x 4 matrices. In the vector form, the product operators, such as $A \otimes I_4$, are 16 x 16 matrices.

We will include one additional complication before writing down our final equation. We will allow a scalar potential in each operator $H_i$:

$$H_i = \vec{\alpha}_i \cdot \vec{p}_i + \beta_i(m_i + S_i(r)). \tag{8}$$

The scalar potential has the same status as the mass – both multiplied by the $\beta$ matrix – whereas the interaction potential $V$ enters the full equation with the same status as the energy $E$, the zero component of a four vector. For this reason, $S$ and $V$ are referred to as scalar and vector potentials.

Thus, the final explicit form of the equation we are considering is:

$$\left[(\vec{\alpha}_1 \cdot \vec{p} + \beta_1(m_1 + S_1))_{\mu\nu}\delta_{\rho\sigma} + \delta_{\mu\nu}(-\vec{\alpha}_2 \cdot \vec{p} + \beta_2(m_2 + S_2))_{\rho\sigma} + V(r)\delta_{\mu\nu}\delta_{\rho\sigma}\right]\Psi_{\nu\sigma} = E\Psi_{\mu\rho} \tag{9}$$

As an example, conider the $\mu = 1, \rho = 1$ component of this equation:

$$\left[(\vec{\alpha}_1 \cdot \vec{p} + \beta_1(m_1 + S_1))_{1\nu}\delta_{1\sigma} + \delta_{1\nu}(-\vec{\alpha}_2 \cdot \vec{p} + \beta_2)_{1\sigma} + V\delta_{1\nu}\delta_{1\sigma}\right]\Psi_{\nu\sigma} = E\Psi_{11},$$

or, using the Kronecker deltas,

$$(\vec{\alpha}_1 \cdot \vec{p} + \beta_1(m_1 + S_1))_{1\nu}\Psi_{\nu 1} + (-\vec{\alpha}_2 \cdot \vec{p} + \beta_2(m_2 + S_2))_{1\sigma}\Psi_{1\sigma} + V\Psi_{11} = E\Psi_{11}. \tag{10}$$

We now use the Dirac representation of the alpha and beta matrices:

$$\vec{\alpha} = \begin{pmatrix} 0 & \vec{\sigma} \\ \vec{\sigma} & 0 \end{pmatrix} \qquad \beta = \begin{pmatrix} 1 & 0 \\ 0 & -1 \end{pmatrix},$$

where each component of these matrics is itself a 2 x 2 matrix, either plus or minus the 2 x 2 unit matrix, or one of the Pauli matrices:

$$\sigma_x = \begin{pmatrix} 0 & 1 \\ 1 & 0 \end{pmatrix}, \qquad \sigma_y = \begin{pmatrix} 0 & -i \\ i & 0 \end{pmatrix}, \qquad \sigma_z = \begin{pmatrix} 1 & 0 \\ 0 & -1 \end{pmatrix}.$$

We can then write out the first term on the left side of Eq. (10):

$$(m_1 + S_1)\Psi_{11} + p_z\Psi_{31} + (p_x - ip_y)\Psi_{41}$$

and recognize that we can also write this as

$$(m_1 + S_1)\Psi_{11} + \left[\begin{pmatrix} p_z & p_x - ip_y \\ p_x + ip_y & -p_z \end{pmatrix}\begin{pmatrix} \Psi_{31} \\ \Psi_{41} \end{pmatrix}\right]_1 = (m_1 + S_1)\Psi_{11} + \left[\vec{\sigma}\cdot\vec{p}\begin{pmatrix} \Psi_{31} \\ \Psi_{41} \end{pmatrix}\right]_1, \quad (11)$$

where the subscript on the square parentheses indicates the we should take the one- (i.e., top-) component of the enclosed matrix multiplication. Similarly, we can write out the second term of Eq. (10):

$$(m_2 + S_2)\Psi_{11} - p_z\Psi_{13} - (p_x - ip_y)\Psi_{14}$$

and recognize this as

$$(m_2 + S_2)\Psi_{11} - \left[\vec{\sigma}\cdot\vec{p}\begin{pmatrix} \Psi_{13} \\ \Psi_{14} \end{pmatrix}\right]_1.$$

Thus Eq. (10) can be written

$$(m_1 + S_1 + m_2 + S_2)\Psi_{11} + \left[\vec{\sigma}\cdot\vec{p}\begin{pmatrix} \Psi_{31} \\ \Psi_{41} \end{pmatrix}\right]_1 - \left[\vec{\sigma}\cdot\vec{p}\begin{pmatrix} \Psi_{13} \\ \Psi_{14} \end{pmatrix}\right]_1 + V\Psi_{11} = E\Psi_{11}. \quad (12)$$

Proceeding in this way for each of the components of Ψ, we have 16 similar equations, which we can write as follows:

Upper-upper

$$\left[\vec{\sigma}_1\cdot\vec{p}\begin{pmatrix} \Psi_{31} \\ \Psi_{41} \end{pmatrix}\right]_1 - \left[\vec{\sigma}_2\cdot\vec{p}\begin{pmatrix} \Psi_{13} \\ \Psi_{14} \end{pmatrix}\right]_1 + V\Psi_{11} = (E - m_1 - S_1 - m_2 - S_2)\Psi_{11} \quad (13a)$$

$$\left[\vec{\sigma}_1\cdot\vec{p}\begin{pmatrix} \Psi_{32} \\ \Psi_{42} \end{pmatrix}\right]_1 - \left[\vec{\sigma}_2\cdot\vec{p}\begin{pmatrix} \Psi_{13} \\ \Psi_{14} \end{pmatrix}\right]_2 + V\Psi_{12} = (E - m_1 - S_1 - m_2 - S_2)\Psi_{12} \quad (13b)$$

$$\left[\vec{\sigma}_1\cdot\vec{p}\begin{pmatrix} \Psi_{31} \\ \Psi_{41} \end{pmatrix}\right]_2 - \left[\vec{\sigma}_2\cdot\vec{p}\begin{pmatrix} \Psi_{23} \\ \Psi_{24} \end{pmatrix}\right]_1 + V\Psi_{21} = (E - m_1 - S_1 - m_2 - S_2)\Psi_{21} \quad (13c)$$

$$\left[\vec{\sigma}_1\cdot\vec{p}\begin{pmatrix} \Psi_{32} \\ \Psi_{42} \end{pmatrix}\right]_2 - \left[\vec{\sigma}_2\cdot\vec{p}\begin{pmatrix} \Psi_{23} \\ \Psi_{24} \end{pmatrix}\right]_2 + V\Psi_{22} = (E - m_1 - S_1 - m_2 - S_2)\Psi_{22} \quad (13d)$$

Upper-lower

$$\left[\vec{\sigma}_1\cdot\vec{p}\begin{pmatrix} \Psi_{33} \\ \Psi_{43} \end{pmatrix}\right]_1 - \left[\vec{\sigma}_2\cdot\vec{p}\begin{pmatrix} \Psi_{11} \\ \Psi_{12} \end{pmatrix}\right]_1 + V\Psi_{13} = (E - m_1 - S_1 + m_2 + S_2)\Psi_{13} \quad (13e)$$

$$\left[\vec{\sigma}_1\cdot\vec{p}\begin{pmatrix} \Psi_{34} \\ \Psi_{44} \end{pmatrix}\right]_1 - \left[\vec{\sigma}_2\cdot\vec{p}\begin{pmatrix} \Psi_{11} \\ \Psi_{12} \end{pmatrix}\right]_2 + V\Psi_{14} = (E - m_1 - S_1 + m_2 + S_2)\Psi_{14} \quad (13f)$$

$$\left[\vec{\sigma}_1\cdot\vec{p}\begin{pmatrix} \Psi_{33} \\ \Psi_{43} \end{pmatrix}\right]_2 - \left[\vec{\sigma}_2\cdot\vec{p}\begin{pmatrix} \Psi_{21} \\ \Psi_{22} \end{pmatrix}\right]_1 + V\Psi_{23} = (E - m_1 - S_1 + m_2 + S_2)\Psi_{23} \quad (13g)$$

$$\left[\vec{\sigma}_1\cdot\vec{p}\begin{pmatrix} \Psi_{34} \\ \Psi_{44} \end{pmatrix}\right]_2 - \left[\vec{\sigma}_2\cdot\vec{p}\begin{pmatrix} \Psi_{21} \\ \Psi_{22} \end{pmatrix}\right]_2 + V\Psi_{24} = (E - m_1 - S_1 + m_2 + S_2)\Psi_{24} \quad (13h)$$

Lower-upper

$$\left[\vec{\sigma}_1 \cdot \vec{p} \begin{pmatrix} \Psi_{11} \\ \Psi_{21} \end{pmatrix}\right]_1 - \left[\vec{\sigma}_2 \cdot \vec{p} \begin{pmatrix} \Psi_{33} \\ \Psi_{34} \end{pmatrix}\right]_1 + V\Psi_{31} = (E + m_1 + S_1 - m_2 - S_2)\Psi_{31} \qquad (13i)$$

$$\left[\vec{\sigma}_1 \cdot \vec{p} \begin{pmatrix} \Psi_{12} \\ \Psi_{22} \end{pmatrix}\right]_1 - \left[\vec{\sigma}_2 \cdot \vec{p} \begin{pmatrix} \Psi_{33} \\ \Psi_{34} \end{pmatrix}\right]_2 + V\Psi_{32} = (E + m_1 + S_1 - m_2 - S_2)\Psi_{32} \qquad (13j)$$

$$\left[\vec{\sigma}_1 \cdot \vec{p} \begin{pmatrix} \Psi_{11} \\ \Psi_{21} \end{pmatrix}\right]_2 - \left[\vec{\sigma}_2 \cdot \vec{p} \begin{pmatrix} \Psi_{43} \\ \Psi_{44} \end{pmatrix}\right]_1 + V\Psi_{41} = (E + m_1 + S_1 - m_2 - S_2)\Psi_{41} \qquad (13k)$$

$$\left[\vec{\sigma}_1 \cdot \vec{p} \begin{pmatrix} \Psi_{12} \\ \Psi_{22} \end{pmatrix}\right]_2 - \left[\vec{\sigma}_2 \cdot \vec{p} \begin{pmatrix} \Psi_{43} \\ \Psi_{44} \end{pmatrix}\right]_2 + V\Psi_{42} = (E + m_1 + S_1 - m_2 - S_2)\Psi_{42} \qquad (13l)$$

Lower-lower

$$\left[\vec{\sigma}_1 \cdot \vec{p} \begin{pmatrix} \Psi_{13} \\ \Psi_{23} \end{pmatrix}\right]_1 - \left[\vec{\sigma}_2 \cdot \vec{p} \begin{pmatrix} \Psi_{31} \\ \Psi_{32} \end{pmatrix}\right]_1 + V\Psi_{33} = (E + m_1 + S_1 + m_2 + S_2)\Psi_{33} \qquad (13m)$$

$$\left[\vec{\sigma}_1 \cdot \vec{p} \begin{pmatrix} \Psi_{14} \\ \Psi_{24} \end{pmatrix}\right]_1 - \left[\vec{\sigma}_2 \cdot \vec{p} \begin{pmatrix} \Psi_{31} \\ \Psi_{32} \end{pmatrix}\right]_2 + V\Psi_{34} = (E + m_1 + S_1 + m_2 + S_2)\Psi_{34} \qquad (13n)$$

$$\left[\vec{\sigma}_1 \cdot \vec{p} \begin{pmatrix} \Psi_{13} \\ \Psi_{23} \end{pmatrix}\right]_2 - \left[\vec{\sigma}_2 \cdot \vec{p} \begin{pmatrix} \Psi_{41} \\ \Psi_{42} \end{pmatrix}\right]_1 + V\Psi_{43} = (E + m_1 + S_1 + m_2 + S_2)\Psi_{43} \qquad (13o)$$

$$\left[\vec{\sigma}_1 \cdot \vec{p} \begin{pmatrix} \Psi_{14} \\ \Psi_{24} \end{pmatrix}\right]_2 - \left[\vec{\sigma}_2 \cdot \vec{p} \begin{pmatrix} \Psi_{41} \\ \Psi_{42} \end{pmatrix}\right]_2 + V\Psi_{44} = (E + m_1 + S_1 + m_2 + S_2)\Psi_{44} \qquad (13p)$$

In these equations, $\vec{\sigma}_1$ and $\vec{\sigma}_2$ are the same standard 2 x 2 Pauli matrix; the subscripts are just a reminder that they are acting on particle 1 and 2, respectively. Note that, in the above equations, $\vec{\sigma}_1$ only acts on components along a column (same second subscript of $\Psi_{\mu\nu}$) and $\vec{\sigma}_2$ only acts on components along a row (same first subscript). This would be the appropriate action on particle 1 and particle 2 respectively (recall the form of $\Psi_{\mu\nu}$ in the product case, Eq. (5)). The $\vec{\sigma} \cdot \vec{p}$ matrices in these equations act on the indicated components of the matrix wave function written as a two-component vector; the subscript on the square bracket denotes the component of the corresponding vector to be taken in each case.

Now we want to consider special cases. We are interested in quarkonium, the bound state of a quark and its antiquark. So $m_1 = m_2$ and we will take $S_1 = S_2 \equiv S/2$. The lowest lying state should be the pseudoscalar $J^P = 0^-$ (in spectroscopic notation, the $^1S_0$ state – spin 0, orbital angular momentum 0, total angular momentum 0). The spin-angular form of the eigenvector matrix is

$$\Psi = \begin{pmatrix} 0 & f & \sin\theta e^{-i\varphi} g_1 & -\cos\theta g_1 \\ -f & 0 & -\cos\theta g_1 & -\sin\theta e^{i\varphi} g_1 \\ \sin\theta e^{-i\varphi} g_2 & -\cos\theta g_2 & 0 & h \\ -\cos\theta g_2 & -\sin\theta e^{i\varphi} g_2 & -h & 0 \end{pmatrix}, \qquad (14)$$

where $f$, $g_1$, $g_2$, and $h$ are functions of $r$. The top left and bottom right quadrants are spin-singlet states written in matrix form; the upper right and lower left quadrants correspond to spin-triplet and P-wave orbital angular momentum coupled to $J = 0$. This should be a state of definite parity and it is. The parity operation on the wave function is

$$\Psi' = \beta_1 \Psi(-\vec{r}) \beta_2, \tag{15}$$

using Eq. (7). Here, the two $\beta$ matrices are the same Dirac $\beta$ matrix; the subscripts just indicate the particle each is acting on. So, if

$$\Psi = \begin{pmatrix} \Psi_a & \Psi_b \\ \Psi_c & \Psi_d \end{pmatrix}, \tag{16}$$

then

$$\Psi' = \begin{pmatrix} \Psi_a(-\vec{r}) & -\Psi_b(-\vec{r}) \\ -\Psi_c(-\vec{r}) & \Psi_d(-\vec{r}) \end{pmatrix}. \tag{17}$$

Since the upper-left and lower-right quadrants are S states and the upper-right and lower-left quadrants are P states, the parity of the spatial-spin wave function (Eq. 17) is positive. Then, since the intrinsic parities of fermions and anti-fermions are opposite, the parity of our state is negative, as it should be.

Returning to the component equations (13) above, we can then use

$$\vec{\sigma} \cdot \vec{p} = -i\vec{\sigma} \cdot \nabla = -i\vec{\sigma} \cdot \left( \hat{r} \frac{\partial}{\partial r} + \hat{\theta} \frac{1}{r} \frac{\partial}{\partial \theta} + \hat{\varphi} \frac{1}{r \sin\theta} \frac{\partial}{\partial \varphi} \right) \tag{18}$$

and, with the expressions for the spherical unit vectors in terms of the Cartesian ones and the standard forms of the Pauli matrices $\sigma_x, \sigma_y, \sigma_z$, this becomes

$$\vec{\sigma} \cdot \vec{p} = -i \begin{pmatrix} \cos\theta & \sin\theta e^{-i\varphi} \\ \sin\theta e^{i\varphi} & -\cos\theta \end{pmatrix} \frac{\partial}{\partial r}$$

$$-\frac{i}{r} \begin{pmatrix} -\sin\theta & \cos\theta e^{-i\varphi} \\ \cos\theta e^{i\varphi} & \sin\theta \end{pmatrix} \frac{\partial}{\partial \theta} - \frac{i}{r\sin\theta} \begin{pmatrix} 0 & -ie^{-i\varphi} \\ ie^{i\varphi} & 0 \end{pmatrix} \frac{\partial}{\partial \varphi}. \tag{19}$$

We can now use this expression and the spin-angular form of the matrix eigenvector given above to obtain the equations for the radial functions:

$$i\frac{d}{dr}(g_1 + g_2) + \frac{2i}{r}(g_1 + g_2) + Vf = (E - m_1 - m_2 - S)f \tag{20a}$$

$$i\frac{df}{dr} + i\frac{dh}{dr} + Vg_1 = (E - m_1 + m_2)g_1 \tag{20b}$$

$$i\frac{df}{dr} + i\frac{dh}{dr} + Vg_2 = (E + m_1 - m_2)g_2 \tag{20c}$$

$$i\frac{d}{dr}(g_1 + g_2) + \frac{2i}{r}(g_1 + g_2) + Vh = (E + m_1 + m_2 + S)h \tag{20d}$$

As an illustration, we will provide the details of this for one case. Consider the 12-component

$$\left[\vec{\sigma}\cdot\vec{p}\begin{pmatrix}\Psi_{32}\\\Psi_{42}\end{pmatrix}\right]_1 - \left[\vec{\sigma}\cdot\vec{p}\begin{pmatrix}\Psi_{13}\\\Psi_{14}\end{pmatrix}\right]_2 + V\Psi_{12} = (E - m_1 - m_2 - S)\Psi_{12}. \tag{21}$$

Using the form of $\vec{\sigma}\cdot\vec{p}$ in Eq. (19) and the components of $\Psi$ in Eq. (14),

$$\vec{\sigma}\cdot\vec{p}\begin{pmatrix}\Psi_{32}\\\Psi_{42}\end{pmatrix} = \vec{\sigma}\cdot\vec{p}\begin{pmatrix}-\cos\theta\, g_2(r)\\-\sin\theta\, e^{i\varphi} g_2(r)\end{pmatrix} = -i\begin{pmatrix}\cos\theta & \sin\theta\, e^{-i\varphi}\\\sin\theta\, e^{i\varphi} & -\cos\theta\end{pmatrix}\begin{pmatrix}-\cos\theta\\-\sin\theta\, e^{i\varphi}\end{pmatrix}\frac{dg_2}{dr}$$

$$-\frac{i}{r}\begin{pmatrix}-\sin\theta & \cos\theta\, e^{-i\varphi}\\\cos\theta\, e^{i\varphi} & \sin\theta\end{pmatrix}\frac{\partial}{\partial\theta}\begin{pmatrix}-\cos\theta\\-\sin\theta\, e^{i\varphi}\end{pmatrix}g_2$$

$$-\frac{i}{r\sin\theta}\begin{pmatrix}0 & -ie^{-i\varphi}\\ie^{i\varphi} & 0\end{pmatrix}\frac{\partial}{\partial\varphi}\begin{pmatrix}-\cos\theta\\-\sin\theta\, e^{i\varphi}\end{pmatrix}g_2$$

$$= \begin{pmatrix}i\\0\end{pmatrix}\frac{dg_2}{dr} + \begin{pmatrix}2i/r\\0\end{pmatrix}g_2. \tag{22}$$

Similarly,

$$\vec{\sigma}\cdot\vec{p}\begin{pmatrix}\Psi_{13}\\\Psi_{14}\end{pmatrix} = \vec{\sigma}\cdot\vec{p}\begin{pmatrix}\sin\theta\, e^{-i\varphi} g_1(r)\\-\cos\theta\, g_1(r)\end{pmatrix} = \begin{pmatrix}0\\-i\end{pmatrix}\frac{dg_1}{dr} + \begin{pmatrix}0\\-2i/r\end{pmatrix}g_1. \tag{23}$$

Substituting these results in the equation for the 12-component of $\Psi$, Eq. (21), and using $\Psi_{12} = f(r)$ gives

$$i\frac{dg_2}{dr} + \frac{2i}{r}g_2 + i\frac{dg_1}{dr} + \frac{2i}{r}g_1 + Vf = (E - m_1 - m_2 - S)f, \tag{24}$$

in agreement with Eq. (20a). The other equations are derived in a similar way. Some of these equations are simply consistent with the zero values for $\Psi_{11}, \Psi_{22}, \Psi_{33}$, and $\Psi_{44}$; some are repeats of others. Finally, we obtain just the four equations, Eq. (20a) to Eq. (20d).

Let us now specialize to the case of quarkonium. For a quark-antiquark state, $m_1 = m_2 \equiv m$ and, since Eq. (20b) and Eq. (20c) are then the same equation, we can take

$$g_1 = g_2 \equiv g \tag{25}$$

and, from comparing Eq. (20a) and Eq. (20d), we can take

$$(E - 2m - S - V)f = (E + 2m + S - V)h. \tag{26}$$

So, we are left with only two equations:

$$2i\frac{dg}{dr} + \frac{4i}{r}g + Vf = (E - 2m - S)f \tag{27a}$$

$$i\frac{df}{dr} + i\frac{d}{dr}(\lambda f) + Vg = Eg \tag{27b}$$

where we have taken

$$h = \lambda f, \qquad \lambda \equiv \left(\frac{E - 2m - S - V}{E + 2m + S - V}\right)f. \tag{28}$$

To make the equations real, we now take $g \to ig$ and we also use reduced radial wave functions $u$ and $v$, defined by $f = u/r$, $g = v/r$. This gives

$$-2\frac{dv}{dr} - 2\frac{v}{r} + Vu = (E - 2m - S)u \tag{29a}$$

$$(1 + \lambda)\frac{du}{dr} - (1 + \lambda)\frac{u}{r} + \left(\frac{d\lambda}{dr}\right)u + Vv = Ev. \tag{29b}$$

We can write this as

$$\frac{dv}{dr} = -\frac{v}{r} - \frac{1}{2}(E - 2m - V - S)u \tag{30a}$$

$$\frac{du}{dr} = \frac{u}{r} + A_1 u + A_2 v, \tag{30b}$$

where

$$A_1 = -\frac{1}{1 + \lambda}\frac{d\lambda}{dr} \qquad \text{and} \qquad A_2 = \frac{E - V}{1 + \lambda}. \tag{31}$$

With $\lambda$ as defined in Eq. (28),

$$A_1 = \frac{\frac{dS}{dr}(E - V) + \frac{dV}{dr}(2m + S)}{(E + 2m + S - V)(E - V)} \tag{32a}$$

$$A_2 = \frac{1}{2}(E + 2m + S - V). \tag{32b}$$

## 3. Application to charmonium

We will now apply this to the $0^-$ state of charmonium, $\eta_c$. We will use the general form of the Cornell potential[12], but we will choose the vector potential to be

$$V = -\frac{4}{3}\frac{\alpha_s}{r} + \beta_V r \tag{33}$$

and the scalar potential to be

$$S = \beta_S r. \tag{34}$$

This will allow us to adjust the vector-scalar mix in the linear potential. With the following choice of parameters (note: although we are writing $\alpha_s$ to mean the strong coupling constant, we are not committed to a particular predetermined value for this constant, but only feel obligated to choose a value within an acceptable range)

$$m_c = 1.311 \text{ GeV} \qquad \alpha_s = 0.42 \qquad \beta_S = \beta_V = 0.08 \text{ GeV}^2,$$

the eigenvalues of Eqs. (30a) and (30b) with $A_1$ and $A_2$ given by Eqs. (32a) and (32b) are obtained by the shooting method, with the equations solved for a given energy using the Runge-Kutta 4$^{th}$ order method. We find the lowest-lying $0^-$ charmonium bound state at 2.9835 GeV. We identify this as the $\eta_c$ with the experimental mass 2983.9 ± 0.5 MeV (Particle Data Group, 2020). With the same parameters, we find a bound state with one node at 3.523 GeV. We identify this as the $\eta_c'$ with an experimental mass 3637.5 ± 1.1 MeV (same source). This agreement is quite rough for $\eta_c'$ (3.5 vs. 3.6 GeV) but it was not our attention to simply adjust the model to fit these two states. If it were, we could certainly do a better job. But this would be rather meaningless; fitting two data points with four parameters is not an achievement. We chose a reasonable set of parameters to produce the $\eta_c$ mass and are content that we then obtained another $0^-$ bound state below charm threshold in approximately the right location. This is an encouraging test of the overall procedure outlined in this paper.

Once this has been established, our main interest is investigating the effect on the bound state of the vector-scalar mix of the potential. To do this we wish to focus on one state, the $\eta_c$, and chose (rather arbitrarily) a fit to the observed mass at the 50-50 mix. We will explore this further in the next section.

## 4. Behavior of the wave function at small and large r

At small $r$, the constant $\lambda$ defined in Eq. (28) goes to 1 (the value unity) and $d\lambda/dr \to 4m/\alpha$. Then Eqs. (30a) and (30b) reduce to

$$-2\frac{dv}{dr} - 2\frac{v}{r} - \frac{\alpha}{r}u = (E - 2m)u \tag{35a}$$

$$2\frac{du}{dr} - 2\frac{u}{r} + \frac{4m}{\alpha}u - \frac{\alpha}{r}v = Ev, \tag{35b}$$

where we have now defined $\alpha = 4\alpha_s/3$. We now take as a trial small-$r$ solution

$$u = Ar^\gamma \quad \text{and} \quad v = Br^\gamma, \tag{36}$$

where $A$ and $B$ are constants. Substituting these in Eqs. (35a) and (35b) gives

$$-2\gamma Br^{\gamma-1} - 2Br^{\gamma-1} - \alpha Ar^{\gamma-1} = (E - 2m)Ar^\gamma \tag{37a}$$

$$2\gamma Ar^{\gamma-1} - 2Ar^{\gamma-1} + \frac{4m}{\alpha}Ar^\gamma - \alpha Br^{\gamma-1} = EBr^\gamma. \tag{37b}$$

Since we require $\gamma > 0$ (for vanishing $u$ and $v$ at the origin), these become in the limit of small $r$,

$$-2\gamma B - 2B - \alpha A = 0 \tag{38a}$$
$$2\gamma A - 2A - \alpha B = 0. \tag{38b}$$

The condition for a nontrivial solution then gives (requiring again a positive value)

$$\gamma = \sqrt{1 - \alpha^2/4}. \tag{39}$$

The wave functions $f$ and $g$ thus both behave as $r^{\gamma-1}$ at the origin. This is reminiscent of the behavior of the wave function in the one-body Dirac Coulomb problem. The change from $\alpha^2$ to $\alpha^2/4$ is expected from considering the nonrelativistic limit of the two-body problem. As in the one-body case the wave function has a mild singularity at the origin.

Let us return now to the mixture of scalar and vector components in the confining potential. The bound-state energies for different choices of $\beta_S$ and $\beta_V$ and the same value of their sum is presented in Table 1 below. In each case, the energy is obtained by the method described above (shooting method with 4th order Runge-Kutta).

The variation of the bound-state energy is modest as we vary the mix from 100% vector to 100% scalar. However, this variation may be useful if we apply this model to the rest of the charmonium spectrum, particularly the P (the charmonium $\chi$) states.

| $\beta_S$ (GeV²) | $\beta_V$ (GeV²) | E (GeV) |
|---|---|---|
| 0 | 0.16 | 3.0103 |
| 0.04 | 0.12 | 2.9964 |
| 0.08 | 0.08 | 2.9835 |
| 0.12 | 0.04 | 2.9715 |
| 0.16 | 0 | 2.9603 |

Table 1

The real issue arises if we consider the large-$r$ behavior of the states. In this limit, it is straightforward to show that the constants $A_1$ and $A_2$ in Eqs. (32a) and (32b) with $V$ and $S$ given by Eqs. (33) and (34) approach the limits

$$A_1 \to 0 \quad \text{and} \quad A_2 \to \frac{1}{2}(\beta_S - \beta_V)r. \tag{40}$$

(For $A_1$, we must consider the cases $\beta_S \neq \beta_V$ and $\beta_S = \beta_V$ separately, but in both cases we conclude that $A_1$ goes to zero in this limit.) So, as $r \to \infty$ Eqs. (30a) and (30b) reduce to

$$\frac{dv}{dr} = \frac{1}{2}(\beta_S + \beta_V)ru \tag{41a}$$

$$\frac{du}{dr} = \frac{1}{2}(\beta_S - \beta_V)rv. \tag{41b}$$

Then

$$\frac{d^2u}{dr^2} = \frac{1}{2}(\beta_S - \beta_V)v + \frac{1}{2}(\beta_S - \beta_V)r\frac{dv}{dr}$$

$$= \frac{1}{2}(\beta_S - \beta_V)\frac{1}{\frac{1}{2}(\beta_S - \beta_V)r}\frac{du}{dr} + \frac{1}{2}(\beta_S - \beta_V)r\frac{1}{2}(\beta_S + \beta_V)ru$$

$$= \frac{1}{r}\frac{du}{dr} + \frac{1}{4}(\beta_S^2 - \beta_V^2)r^2 u. \tag{42}$$

So, in the limit of infinite $r$, if $\beta_S \neq \beta_V$, the first term is dominated by the $r^2$-term and we have the equation

$$\frac{d^2u}{dr^2} = \frac{1}{4}(\beta_S^2 - \beta_V^2)r^2 u. \tag{43}$$

The solution of this, in the same limit, is

$$u = A\exp\left(\pm\sqrt{\frac{1}{16}(\beta_S^2 - \beta_V^2)}\,r^2\right). \tag{44}$$

Now, if $\beta_S > \beta_V$, we can take

$$\frac{1}{16}(\beta_S^2 - \beta_V^2) = k^2, \tag{45}$$

with $k$ real and taken to be positive. Then, in the large-$r$ limit,

$$u \to Ae^{\pm kr^2}. \tag{46}$$

But we must choose the minus sign for finiteness. Therefore, if $\beta_S > \beta_V$, the wave function behaves at large $r$ as $e^{-kr^2}$, appropriate for a bound state. On the other hand, if $\beta_V > \beta_S$, we can take

$$\frac{1}{16}(\beta_S^2 - \beta_V^2) = -k^2 \tag{47}$$

with $k$ real. Then the behavior at large $r$ will be

$$u \to Ae^{\pm\sqrt{-k^2}\,r^2} = Ae^{\pm ikr^2} \tag{48}$$

with $k$ real, or more appropriately for real $u$:

$$u \to A\cos kr^2 + B\sin kr^2. \tag{49}$$

This is oscillatory and is not characteristic of a bound state.

We still must consider the case $\beta_S = \beta_V \equiv \beta$. In this case, Eqs. (40a) and (40b) reduce to

$$\frac{dv}{dr} = \beta r u \tag{50a}$$

$$\frac{du}{dr} = 0. \tag{50b}$$

So we have $u = \text{const.} \equiv A$ and then $v = \frac{1}{2}\beta r^2 A$. Then we must choose $A = 0$ for finiteness. Strictly, the solution of Eq. (50a) could include a constant, but this would give a wave function $g = v/r$ going as constant/$r$ at large $r$ and we must then require that the constant be zero for normalizability. Therefore, the behavior at large $r$ in this case is that of a bound state.

We note that in the cases $\beta_S \geq \beta_V$ we can use behavior at infinity to eliminate unwanted solutions, but that is not the case for Eq. (49); there is no argument for taking $A = B = 0$ in this limit. We conclude that oscillatory behavior at large $r$ is allowed if $\beta_V > \beta_S$ and therefore that the states of this type in Table 1 above are not true bound states; they are only quasibound. For true bound states, we thus require $\beta_S \geq \beta_V$.

Is there an explanation for oscillatory behavior at large $r$ in the case of a vector-dominated confining potential? The two-body Dirac equation for equal-mass particles has a negative energy continuum starting at zero total energy (one particle at energy $m$, the other at energy $-m$). But the linear potential will affect the level of this continuum. (We will ignore the short-range potential in this discussion of the effect of the long-range potential.) The line of the negative-energy continuum is brought down by the scalar part of the linear potential and raised by the vector part. If $\beta_S \geq \beta_V$, this line is either unaffected or brought down. On the other hand, if $\beta_V > \beta_S$ this line will rise as $0 + (\beta_V - \beta_S)r$. Eventually it will reach the level of a given positive-energy state. At that point the positive-energy state will mix with free negative-energy states. Equivalently we can think of this as tunneling from the positive-energy state to the negative-energy continuum. For a given positive-energy state of energy $E$, the classically allowed region or the state is $r \leq r_1$, where

$$2m + (\beta_S + \beta_V)r_1 = E, \quad \text{or} \quad r_1 = \frac{E - 2m}{\beta_S + \beta_V}. \tag{51}$$

(Recall that we are ignoring the short-range potential.) The line of the negative-energy continuum will reach this level at

$$0 + (-\beta_S + \beta_V)r_2 = E, \quad \text{or} \quad r_2 = \frac{E}{\beta_V - \beta_S}. \tag{52}$$

There is thus a classically forbidden region of width $r_2 - r_1$ through which the tunneling occurs. As an example, consider the $\beta_V = 16, \beta_S = 0$ case in Table 1. Then

$$r_1 = \frac{3.0193 \text{ GeV} - 2(1.311 \text{ GeV})}{0.16 \text{ GeV}^2} = 2.4 \text{ GeV}^{-1} \quad \text{and} \quad r_2 = \frac{3.0103 \text{ GeV}}{0.16 \text{ GeV}^2} = 18.8 \text{ GeV}^{-1}.$$

So the classically forbidden region has width 16.4 GeV$^{-1}$. Clearly, the tunneling will be very small and the lifetime of the quasibound state very large on the particle physics timescale. However, since the confinement, under ordinary circumstances, is expected to be absolute, it may be wise for model builders to avoid a vector-dominated linear confinement. For further discussion of the tunneling scenario (but in the one-body Dirac case), see Reference [13].

## 5. First-order corrections

For the Breit equation we are considering (no retardation) some spin dependence is included in the structure of the equation. However, explicit spin dependence in the nonrelativistic limit is lacking. In this approximation we will get some spin-terms, of the form $\vec{L} \cdot \vec{S}_1$ and $\vec{L} \cdot \vec{S}_2$, which in the equal-mass case go over to $\vec{L} \cdot (\vec{S}_1 + \vec{S}_2) = \vec{L} \cdot \vec{S}$. But there is no explicit spin-spin or tensor terms. These only occur in this reduction if we include the Breit retardation term. This term can be included for the short-range interaction in the QCD-inspired model we are considering. This interaction is attributed to single-gluon exchange and a retarded effect of the same form as in the electromagnetic case would be expected. We would then include the term

$$V_B = \frac{2}{3}\frac{\alpha_s}{r}(\vec{\alpha}_1 \cdot \vec{\alpha}_2 + \vec{\alpha}_1 \cdot \hat{r}\vec{\alpha}_2 \cdot \hat{r}) \tag{53}$$

in Eq. (1). The possibility of including a retardation term for the confining potential, particularly a vector piece of this potential, is an open question; we will choose not to do so. For the heavy quarkonia we are considering, charmonium or, even more so, upsilonium, we expect the effect of the retardation terms to be relatively small because of the compact nature of the states. Thus, we would expect that first-order perturbation theory would be adequate to account for this effect. In fact, the use of higher-order perturbation theory with this potential may not be appropriate. As argued in Reference [14], in the electromagnetic case the first-order correction gives the correct result, but higher-order corrections do not. We expect the same situation in the QCD case. So we would calculate the energy shift due to the inclusion of the retardation term to be

$$\Delta E = \langle \Psi^{(0)} | V_B | \Psi^{(0)} \rangle \tag{54}$$

where $\Psi^{(0)}$ is the wave function calculated in Section 3. If we evaluate $V_B$ in the Pauli (nonrelativistic) approximation to the Breit equation, we obtain a kinematical term and spin-dependent terms: spin-orbit, tensor, and spin-spin terms.

The kinematical term is taken care of by the use of the wave function of the full non-retarded Breit equation. Also, for the states treated explicitly in this work (spin singlet s-wave states) neither the spin-orbit nor the tensor interactions will contribute. That leaves only the spin-spin term. This term, in the Pauli approximation is

$$V_{spin} = -\frac{8\pi\alpha_s}{3m^2}\delta(\vec{r}). \tag{55}$$

So the energy shift in this approximation is

$$\Delta E = \langle \Psi^{(0)} | V_{spin} | \Psi^{(0)} \rangle. \tag{56}$$

As noted above, the wave function in this expression has a mild singularity at the origin and this causes an obvious problem for the evaluation of the energy shift. We do not judge this singularity to be physical; it is due to the assumed point nature of the short-range potential in an equation of the Dirac type. One can address this problem by smearing the short-range potential.[15] This would require some additional parametrization in our model. Rather than doing this, we believe that, since the approximation used in deriving the expression for $V_{spin}$ was a nonrelativistic one, a reasonable approximation to this shift may be obtained by using a nonrelativistic wave function, i.e., we will take

$$\Delta E_{spin} \cong \langle \psi | V_{spin} | \psi \rangle, \tag{57}$$

with $\psi$ being the ground state of the Hamiltonian

$$H = \frac{\vec{p}^2}{2\mu} - \frac{4}{3}\frac{\alpha_s}{r} + \beta r, \tag{58}$$

where $\mu$ is the reduced mass, $\mu = m/2$), and we take the same parameters as before:

$$m = 1.311 \text{ GeV}, \alpha_s = 0.42, \beta = \beta_V + \beta_S = 0.18 \text{ GeV}^2.$$

We can diagonalize this Hamiltonian in a traditional way by expanding our wave function in a basis of three-dimensional isotropic oscillator eigenfunctions (radial oscillator functions). This yields a ground state close in energy (with the addition of the mass term) to the value obtained from the instantaneous Breit equation, which is not surprising since the latter reduces to Eq, (58) in the nonrelativistic limit, a limit which should be a good approximation for charmonium. For this energy, we obtain a value of the wave function at the origin of $\psi(0) = 0.1199 \text{ GeV}^{3/2}$. This gives an energy shift

$$\Delta E = -\frac{8\pi\alpha_s}{3(1.311 \text{ GeV})^2}(0.1199 \text{ GeV}^{3/2})^2 = -29.4 \text{ MeV}. \tag{59}$$

So, the correction is possibly significant, but is still only at the 1% level. At least for heavy quarkonium, it is legitimate to treat it as a small correction. For another treatment of this issue, see Reference [16].

## 6. Summary and future work

In this work, the reduction of the instantaneous Breit equation for pseudoscalar states to a system of first-order differential equations for radial wave functions has been elaborated. There are four equations in this system for unequal masses, but these can be reduced to two equations in the equal mass case. This reduction is rather laborious. However, once performed, the use of these equations to explore potential models is straightforward. These equations have been tested by application to the pseudoscalar states of charmonium with the use of a Cornell-type potential. The scalar-vector mix of the confining potential is explored, and we argue that the restriction to scalar-dominated linear potential is favored.

Extension to the $J^P = 1^-$ states of charmonium is ongoing. There will be an additional complication in this case since the tensor potential mixes the S and D states in the upper-upper and lower-lower sectors. The P states, both the spin triplet $\chi$ states and the spin singlet $h_c$ state, are also under consideration. These states should present a strong test of the spin-dependence of the equation. In particular, the spin-orbit and tensor first-order corrections as well as that of spin-spin may be crucial in obtaining the observed ordering for these states.